\documentclass[conference]{IEEEtran}
\IEEEoverridecommandlockouts

\usepackage{cite}
\usepackage{amsmath,amssymb,amsfonts}
\usepackage{algorithmic}
\usepackage{algorithm}
\usepackage{graphicx}
\usepackage{textcomp}
\usepackage{makecell}
\usepackage{xcolor}
\usepackage{subfigure}
\usepackage[caption=false,font=normalsize,labelfont=sf,textfont=sf]{subfig}
\usepackage{booktabs}

\newcommand{\hdel}[1]{}

\def\BibTeX{{\rm B\kern-.05em{\sc i\kern-.025em b}\kern-.08em
    T\kern-.1667em\lower.7ex\hbox{E}\kern-.125emX}}
\begin{document}
\title{Transfer Generative Adversarial Networks (T-GAN)-based Terahertz Channel Modeling\\
}


\author{
	\IEEEauthorblockN{Zhengdong~Hu, Yuanbo~Li, and Chong~Han}
	\IEEEauthorblockA{Terahertz Wireless Communications (TWC) Laboratory, Shanghai Jiao Tong University, China.\\
	Email: \{huzhengdong, yuanbo.li, chong.han\}@sjtu.edu.cn}
\vspace{-1cm}
}


\maketitle

\thispagestyle{empty}
\pdfoptionpdfminorversion=7
\begin{abstract}
Terahertz (THz) communications are envisioned as a promising technology for 6G and beyond wireless systems, providing ultra-broad bandwidth and thus Terabit-per-second (Tbps) data rates. However, as foundation of designing THz communications, channel modeling and characterization are fundamental to scrutinize the potential of the new spectrum. Relied on physical measurements, traditional statistical channel modeling methods suffer from the problem of low accuracy with the assumed certain distributions and empirical parameters. Moreover, it is time-consuming and expensive to acquire extensive channel measurement in the THz band. In this paper, a transfer generative adversarial network (T-GAN) based modeling method is proposed in the THz band, which exploits the advantage of GAN in modeling the complex distribution, and the benefit of transfer learning in transferring the knowledge from a source task to improve generalization about the target task with limited training data. Specifically, to start with, the proposed GAN is pre-trained using the simulated dataset, generated by the standard channel model from 3rd generation partnerships project (3GPP). Furthermore, by transferring the knowledge and fine-tuning the pre-trained GAN, the T-GAN is developed by using the THz measured dataset with a small amount. Experimental results reveal that the distribution of PDPs generated by the proposed T-GAN method shows good agreement with measurement. Moreover, T-GAN achieves good performance in channel modeling, with 9 dB improved root-mean-square error (RMSE) and higher Structure Similarity Index Measure (SSIM), compared with traditional 3GPP method.



\end{abstract}

\section{Introduction}
With the exponential growth of the number of interconnected devices, the sixth generation (6G) is expected to achieve intelligent connections of everything, anywhere, anytime~\cite{akyildiz2022terahertz}, which demands Tbit/s wireless 
data rates. To fulfill the demand, Terahertz (THz) communications gain increasing attention as a vital technology of 6G systems, thanks to the ultra-broad bandwidth ranging from tens of GHz to hundreds of GHz~\cite{tera_ref}. 
The THz band is promising to address the spectrum scarcity and capacity limitations of current wireless systems, and realize long-awaited applications, extending from wireless cognition, localization/positioning, to integrated sensing and communication~\cite{han2022thz}.

To design reliable THz wireless systems, one fundamental challenge lies in developing an accurate channel model to portray the propagation phenomena. Due to the high frequencies, new characteristics occur in the THz band, such as frequency-selective absorption loss and rough-surface scattering. Attribute to these new characteristics, THz channel modeling is required to capture these characteristics. However, traditional statistical channel modeling methods suffer from the problem of low accuracy with the assumed certain distributions and empirical parameters. For example, a geometric based stochastic channel model (GSCM) assumes that the positions of scatters follow certain statistical distributions, such as the uniform distribution within a circle around the transmitters and receivers~\cite{GSCM}. However, the positions of scatters are hard to characterize by certain statistical distributions, making the GSCM not accurate for utilization in the THz band. Moreover, it is time-consuming and costly to acquire extensive channel measurement for THz channel modeling. To this end, an accurate channel modeling method with limited measurement data for the THz band is needed.

Recently, deep learning (DL) is popular and widely applied in wireless communications, such as channel estimation~\cite{ce2,ce3} and channel state information (CSI) feedback~\cite{csi}. Among different kinds of DL methods, the generative adversarial network (GAN) has the advantage of modeling complex distribution accurately without any statistical assumptions, based on which GAN can be utilized to develop channel models. The authors in~\cite{appro} train GAN to approximate the probability distribution functions (PDFs) of stochastic channel response. In~\cite{channel_gan}, GAN is applied to generate synthetic channel samples close to the distribution of real channel samples. The researchers in~\cite{mimo-gan} model the channel with GAN through channel input-output measurements. In~\cite{distribution}, a model-driven GAN-based channel modeling method is developed in intelligent reflecting surface (IRS) aided communication system. These methods achieve good performance in modeling the channel and prove high consistency between the target channel distribution and the generated  channel distribution. However, the GAN based channel modeling method has not been exploited in the THz band. Moreover, it is a challenge to train GAN for channel modeling with the scarce THz channel measurement dataset.

In this paper, a transfer GAN (T-GAN)-based THz channel modeling method is proposed,  which can learn the distribution of power delay profile (PDP) of the THz channel. Moreover, to tackle the challenge of limited channel measurement in the THz band, the transfer learning technique is introduced in T-GAN, which reduces the size requirement of channel dataset for training and enhances the performance of channel modeling, through transferring the knowledge stored in a pre-trained model to a new model~\cite{transfer1,finegan}. Furthermore, the performance of T-GAN in modeling the channel distribution is validated by real measurements~\cite{yuanbo_icc}.

The contributions of this paper are listed as follows.
\begin{itemize}
    \item We propose a T-GAN based THz channel modeling method, in which a GAN is designed to capture the distribution of PDPs of the THz channel, by training on the dataset of PDP samples. 
    \item To tackle the challenge of limited measurement data for THz channel modeling, transfer learning is further exploited by T-GAN, which reduces the size requirement of training dataset, and enhances the performance of GAN, through transferring the knowledge stored in a pre-trained model to a new model.
\end{itemize}

The rest of the sections are organized as follows. Sec.~\ref{sec_proposed} details the proposed T-GAN based channel modeling method. Sec.~\ref{sec_performance} demonstrates the performance of the proposed T-GAN method. The paper is concluded in Sec.~\ref{sec_conclusion}.

\textbf{Notation:} 
$a$ is a scalar. \textbf{a} denotes a vector. 
\textbf{A} represents a matrix. $\mathbb{E}\{\cdot \}$ describes the expectation. $\nabla$ denotes the gradient operation. $\left\|\cdot\right\| $ represent the L2 norm. $\textbf{I}_N$ defines an $N$ dimensional identity matrix. $\mathcal{N}$ denotes the normal distribution.
\section{Transfer GAN (T-GAN) Based Channel Modeling}\label{sec_proposed}
In this section, the channel modeling problem is first formulated into a channel distribution learning problem. Then, the proposed GAN in T-GAN method is elaborated. Finally, T-GAN is presented.



\subsection{Problem Formulation}\label{sec_system}
    The channel impulse response (CIR) can be represented as 
\begin{equation}
    h(\tau) =\sum_{l=0}^{L-1}\alpha_le^{j\phi_l}\delta(\tau-\tau_l),
\end{equation}
where $\tau_l$ denotes the delay of the $l^{th}$ multi-path components (MPCs), $L$ denotes the number of MPC, $\alpha_l$ refers to the path gain and $\phi_l$ represents the phase of the corresponding MPC. To characterize the channel, PDP is an important feature, which indicates the dispersion of power over the time delay, specifically, the received power with respect to the delay in a multi-path channel. It can be extracted from the channel impulse response by
\begin{equation}
    P(\tau) = |h(\tau)|^2,
\end{equation}
Then, the channel modeling problem is exploited by learning the distribution of PDPs denoted by $p_r$, which is difficult to be analytically represented. Instead, the distribution $p_r$ can be captured by generating fake PDP samples with distribution $p_g$, such that the generated distribution $p_g$ of PDPs can match the actual distribution $p_r$.




\subsection{Proposed GAN}
\begin{figure}
    \centering
    \includegraphics[width=0.45\textwidth]{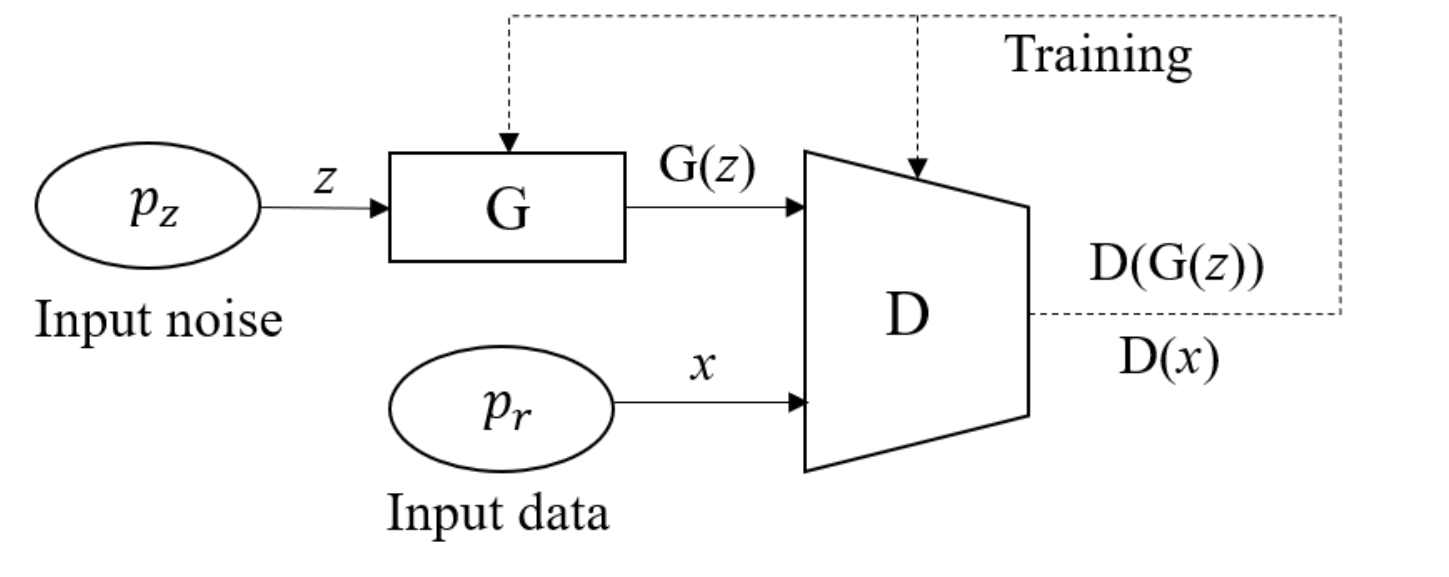}
    \caption{Framework of GAN.}
    \label{fig_gan}
\end{figure}

The GAN can be utilized to learn the distribution of PDPs denoted by $p_r$, with the framework depicted in Fig~\ref{fig_gan}. The GAN consists of two sub-networks, namely, generator and discriminator. The generator is aimed at generating fake samples $G(z)$ to fool the discriminator, in which $z$ is the noise sample, by mapping the input noise distribution $p_z(z)$ to the generated distribution $p_g=p(G(z))$. The discriminator tries to distinguish between real samples $x$ from $p_r$ and fake samples $G(z)$ from $p_g$, and the output of the discriminator $D(x)$ and $D(G(z))$ can be treated as the probability of being a real sample. The two networks are trained in an adversarial manner, which can be considered as a two-player zero-sum minimax game. Specifically, the training objective can be represented by
\begin{equation}
\label{gan_objective}
    \mathop{\min}\limits_{G}\mathop{\max}\limits_{D} \mathbb{E}_{\boldsymbol{x}\sim p_r}[\log D(\boldsymbol{x})]+\mathbb{E}_{\boldsymbol{z}\sim p_z}[\log (1-D(G(\boldsymbol{z})))],
\end{equation}

\noindent where the generator minimizes the probability $(1-D(G(z))$ that the generated sample is detected as fake by the discriminator, while the discriminator maximizes this probability. Therefore, the generator and discriminator compete against each other with the opposite objectives in the training process. Through the adversarial training, the Nash equilibrium can be achieved, such that the generator and discriminator cannot improve their objectives by changing only their own network. Moreover, the global optimum of the training objective can be achieved in the equilibrium when $p_g=p_r$. However, training with the objective function in \eqref{gan_objective} is unstable, since the training objective is potentially not continuous with
respect to the generator’s parameters~\cite{wgan-gp}. Therefore, the improved version of GAN, namely, Wasserstein GAN with gradient penalty (WGAN-GP)~\cite{wgan-gp} is adopted. The modified objective function is expressed as 
\begin{equation}\label{wq_equation}
\begin{aligned}
    \mathop{\min}\limits_{G}\mathop{\max}\limits_{D} \mathbb{E}_{\boldsymbol{x}\sim p_r}[D(\boldsymbol{x})]+&\mathbb{E}_{\boldsymbol{z}\sim p_z}[ (1-D(G(\boldsymbol{z})))]\\
      +&\lambda\mathbb{E}_{\tilde{\boldsymbol{x}}}[(\left\|\nabla_{\tilde{\boldsymbol{x}}}D(\tilde{\boldsymbol{x}})\right\|-1)^2)],
\end{aligned}
\end{equation}
where the last term is the gradient penalty term to enforce Lipschitz constraint that the gradient of the GAN network is upper-bounded by a maximum value, the symbol $\tilde{\boldsymbol{x}}$ is the uniformly sampled point between the points of $\boldsymbol{x}$ and $G(\boldsymbol{z})$. Moreover, the parameter $\lambda$ is the penalty coefficient. 

\begin{figure}[t]
    \centering
    \includegraphics[width=0.5\textwidth]{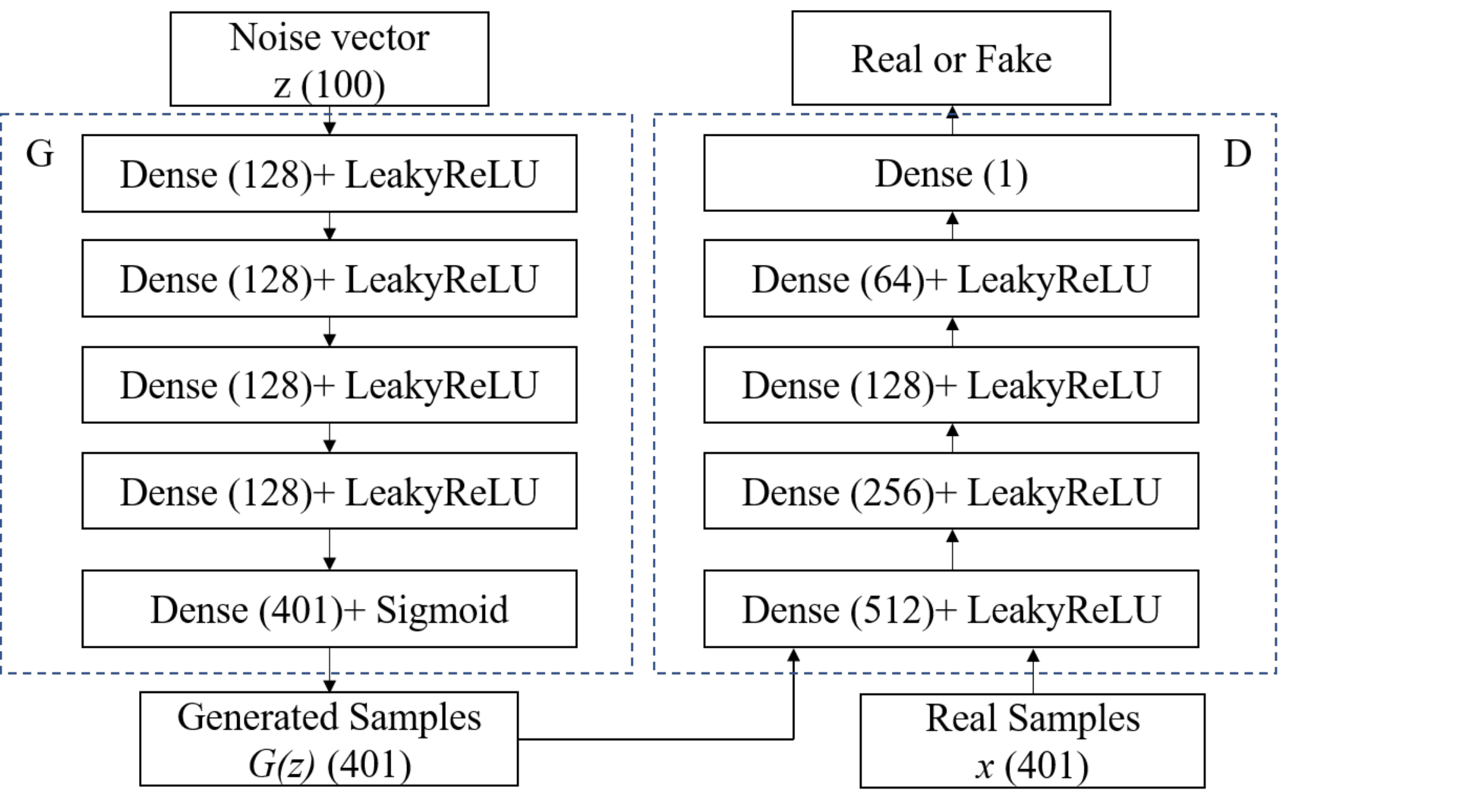}
    \caption{Structure of generator and discriminator.}
    \label{fig_gan_structure}
\end{figure}

After introducing the framework of GAN, the detailed architecture of proposed GAN network is presented. The structures of generator G and discriminator D are depicted in Fig.~\ref{fig_gan_structure}, where the number in the bracket denotes the dimension. The input to the generator is a noise vector $z$ with dimension $n_z=100$, which is sampled from the probability density function $\mathcal{N}(0,\sigma^2\mathbf{I}_{n_z})$. The generator consists of five dense layers, and the numbers of neurons in the dense layers are 128, 128, 128, 128, 401, respectively. It is worth noting that the size of the output layer is equal to the size of PDP. The activation function of the first four dense layers is LeakyReLU function, which can speed up the convergence and avoid the gradient vanishing problem. The formula of the LeakyReLU function is expressed as
\begin{equation}
    f(x) = \begin{cases}
	      x, & \mathrm{if} \ x \geq0 \\
	      \alpha x, & \mathrm{if} \ x <0	
		   \end{cases},
\end{equation}
where $\alpha$ is the slope coefficient when the value of neuron $x$ is negative. In addition to the LeakyReLU function, a Sigmoid function is utilized in the last layer, which maps the output to the range of [0, 1]. The Sigmoid function is defined as 
\begin{equation}
    f(x) = \frac{1}{1+e^{-x}}.
\end{equation}
\noindent After going through the dense layers and activation functions in the generator, the input noise vectors are transformed into the generated samples. Then, the generated samples together with real samples are passed to the discriminator.

The discriminator is designed to distinguish between generated samples and real samples. The numbers of neurons for the five dense layers in the discriminator are 512, 256, 128, 64, 1, respectively. The activation function chosen for the first 4 layers is the LeakyReLU function introduced before. The activation function for the output layer is linear activation function, which is decided by the objective function of WGAN-GP introduced in \eqref{wq_equation}.

\begin{figure}[t]
    \centering
    \includegraphics[width=0.4\textwidth]{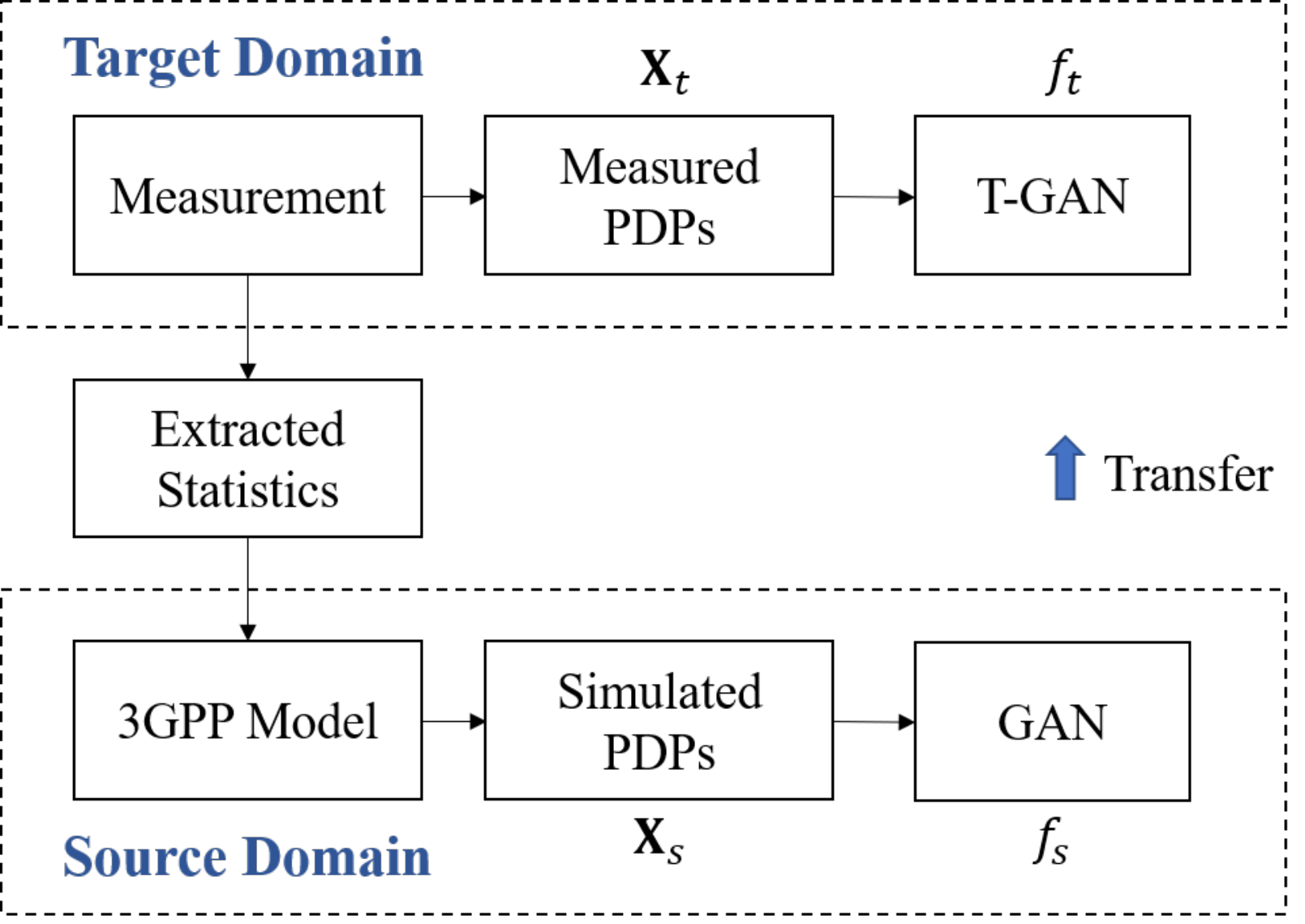}
    \caption{Framework for T-GAN.}
    \label{fig_transfer}
\end{figure}

\subsection{Proposed T-GAN}
The framework for the proposed T-GAN is depicted in Fig.~\ref{fig_transfer}, in which the transfer learning is conducted between the measurement and 3GPP TR 38.901 model~\cite{3gpp}. The measured PDPs denote the PDPs extracted from measurement with a small size, while the simulated PDPs refer to the PDPs simulated by the 3GPP model, which is implemented with the extracted statistics from measurement. Then, the proposed GAN and T-GAN with the same network structure, are trained on the simulated PDPs and measured PDPs, respectively, to capture the distribution of PDPs. Since the size of measured PDPs is quite small for the training of T-GAN, which can cause the difficulty of converging or the over-fitting problem, the transfer learning is exploited to tackle these problems.

To describe the transfer learning formally, a domain denoted by $\mathcal{D}$ consists of a feature space $\mathcal{X}$ and a marginal probability distribution $\mathcal{P}(\mathbf{X})$ defined on $\mathbf{X}=\{\mathbf{x}_1,\mathbf{x}_2,\cdots,\mathbf{x}_N\}\in\mathcal{X}$, where $N$ is the number of feature vectors in $\mathbf{X}$. As depicted in Fig.~\ref{fig_transfer}, the target domain $\mathcal{D}_t$ and source domain $\mathcal{D}_s$ are defined on measurement and 3GPP model, respectively. The feature spaces for the two domains are both constructed by PDPs, with different marginal probability distributions defined on measured PDPs $\mathbf{X_t}$ and simulated PDPs $\mathbf{X_s}$.

Moreover, given a domain $\mathcal{D}(\mathcal{X},P(\mathbf{X}))$, a task denoted by $\mathcal{T}$ is defined by a label space $\mathcal{L}$ and a predictive function $f(\cdot)$, and the predictive function is learned from the pairs $(\mathbf{x}_n,l_n)$ with $\mathbf{x}_n\in \mathbf{X}$ and $l_n\in\mathcal{L}$. In the target domain $\mathcal{D}_t$ and source domain $\mathcal{D}_s$, the tasks are the same to capture the distribution of PDPs, and the label space is $\mathcal{L}=\{0, 1\}$ representing whether the PDP sample is generated by the proposed GAN or from the training dataset. The T-GAN and GAN serve as the predictive functions $f_t$ and $f_s$. Then, transfer learning is aimed at learning the function $f_t$ in target domain $\mathcal{D}_t$ with the knowledge of $\mathcal{T}_s$ in source domain $\mathcal{D}_s$, i.e., transferring the knowledge stored in GAN trained on simulated PDPs to T-GAN trained on the measured PDPs.

The method of fine-tuning~\cite{finegan} is adopted for the transfer learning. The T-GAN is initialized with the weights of the GAN trained on the simulated PDPs, and is then fine-tuned on the measured PDPs with small size. It is worth noting that both the generator and discriminator in the GAN are transferred, which can yield the better performance in generating high quality samples and fast convergence, compared with transferring only the generator or the discriminator~\cite{finegan}.

With transfer learning, the performance of T-GAN can be largely enhanced. Specifically, the channel statistics extracted for 3GPP method are captured by the proposed GAN trained on simulated PDPs, which are further transferred to T-GAN. Moreover, T-GAN can learn the features of PDPs that are not captured by 3GPP method, directly from measurement, which further improves the performance of T-GAN in modeling the distribution of PDPs.

\section{Experiment and Performance Evaluation}\label{sec_performance}
In this section, the experiment settings are elaborated. Moreover, the performance of the T-GAN are evaluated by comparing the generated distribution of PDPs with measurement.

\subsection{Dataset and Setup}
\begin{figure}[t]
    \centering
    \includegraphics[width=0.5\textwidth]{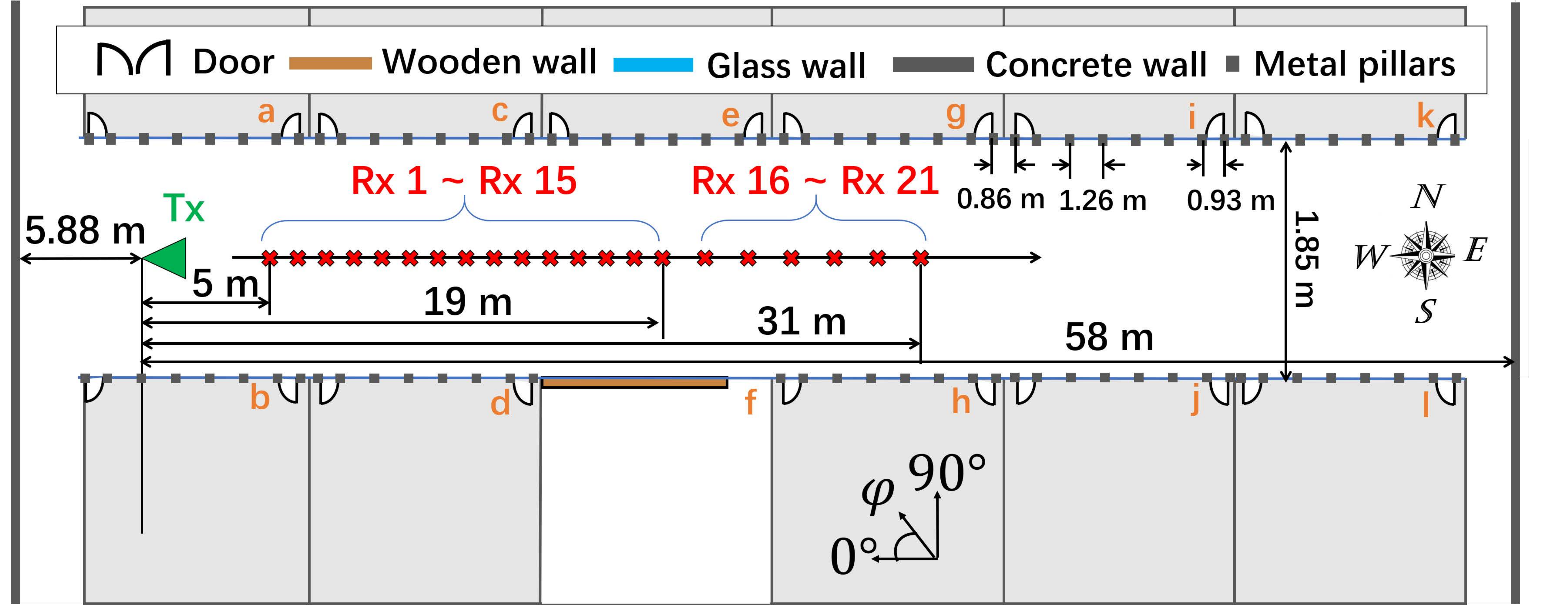}
    \caption{Measurement layout in the indoor corridor scenario~\cite{yuanbo_icc}.}
    \label{fig_scenario}
\end{figure}

The dataset is collected from the measurement campaign in~\cite{yuanbo_icc}. which is conducted in an indoor corridor scenario at 306-321 GHz with 400 ns maximum delay, as depicted in Fig.~\ref{fig_scenario}. With the measurement data, the PDPs can be extracted to characterize the channel in the 21 receiver points. Since the sample frequency interval is relatively small, as 2.5~MHz, the measured PDPs are very long, including 6001 sample points, which results in extraordinary computation and time consumption to train the GANs. To address this problem, we only use the measured channel transfer functions in the frequency band from 314 to 315 GHz, based on which the PDPs can be shorten to 401 sample points.

The PDPs of the 21 measured channels make up the measured dataset. In addition to the measured dataset, the dataset of simulated PDPs can be generated by 3GPP model with the extracted statistics from the measurement, which consists of 10000 channels. Compared to the measured dataset, the simulated dataset has larger data size with the channel statistics embedded. Moreover, the PDPs in two datasets are normalized into the range of [0, 1] by the min-max normalization method.



The training procedure of the GAN network is explained in detail as follows.
Firstly, the input noise vector $z$ of size 100 is generated by the multivariate normal distribution, which can provide the capabilities to transform into the desired distribution. The gradient penalty parameter $\lambda$ in \eqref{wq_equation} is set as 10, which works well in the training process. Moreover, the stochastic gradient descent (SGD) optimizer is applied for the generator network, and the adaptive moment estimation (Adam) optimizer is chosen for the discriminator network. In addition, the learning rates of the two optimizers are both set as 0.0002 to stabilize the training. 




All the experimental results are implemented on a PC with AMD Ryzen Threadripper 3990X @ 2.19 GHz and four Nvidia GeForce RTX 3090 Ti GPUs. In addition, the training of GAN network is carried out in the Pytorch framework.

\subsection{Convergence}
\begin{figure}[t]
    \centering
     \includegraphics[width=0.5\textwidth]{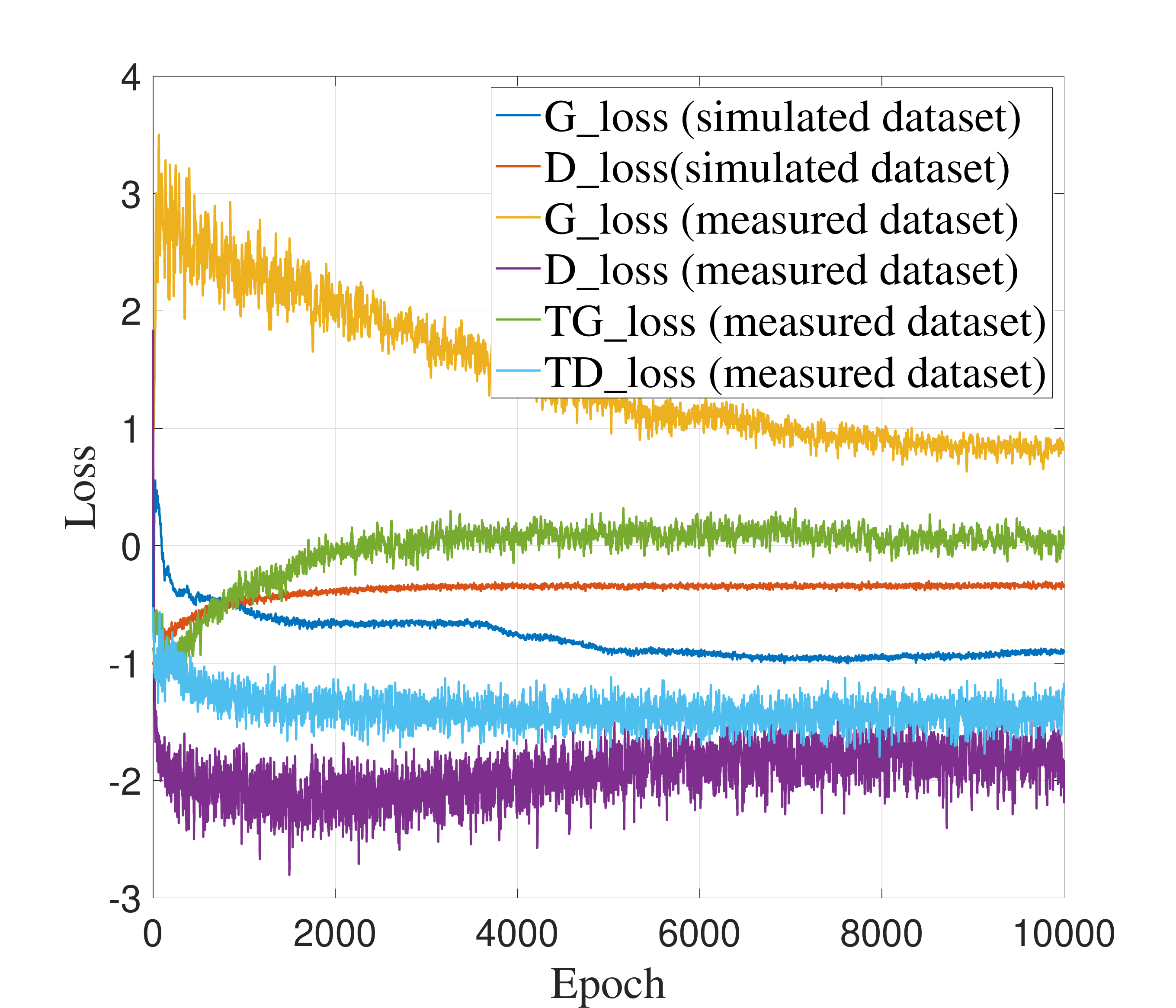}
    \caption{Loss of the generator and discriminator in the GAN network. }
    \label{fig_loss}
\end{figure}

The proposed GAN is first trained on the simulated dataset, and is then fine-tuned on the measured dataset with transfer learning to develop the T-GAN. The numbers of epochs for training the proposed GAN and T-GAN are both set as 10000. A epoch is defined as a complete training cycle through the training dataset, in which the generator and discriminator are iteratively trained for once. To demonstrate the benefits of transfer learning, the GAN is also trained on the measured dataset without transfer learning for comparison. The loss of generator denoted by G\_loss and loss of discriminator denoted by D\_loss are shown in the Fig.~\ref{fig_loss}, in which the TG\_loss and TD\_loss correspond to the losses for T-GAN. For the simulated dataset, it is clear that the generator and discriminator reach the equilibrium in the end. For the measured dataset, the loss of T-GAN is close to the loss for the simulated dataset except for some small fluctuations. The fluctuations are due to the small size of the measured dataset. By comparison, the training is not stable for the GAN network without transfer leaning. There is large fluctuation in the discriminator loss, and the absolute values of G\_loss and D\_loss are quite large compared to the losses for the simulated dataset. The comparison demonstrates the benefits of the transfer learning in the training of GAN network, which enables T-GAN to converge with a small training dataset. Moreover, it takes only 4000 epochs for T-GAN to converge, compared to 6000 epochs for GAN trained on the simulated dataset. The training time of T-GAN on the measured dataset is also small, which is only 114 seconds compared to 7 hours for GAN trained on the simulated dataset. From these results, it is clear that the transfer learning technique can improve the convergence rate of T-GAN,  and reduce the training overhead with the knowledge from the pre-trained model.
\begin{figure*}[t]
    \centering
    \subfigure[Samples of PDP.]{\includegraphics[width=0.45\textwidth]{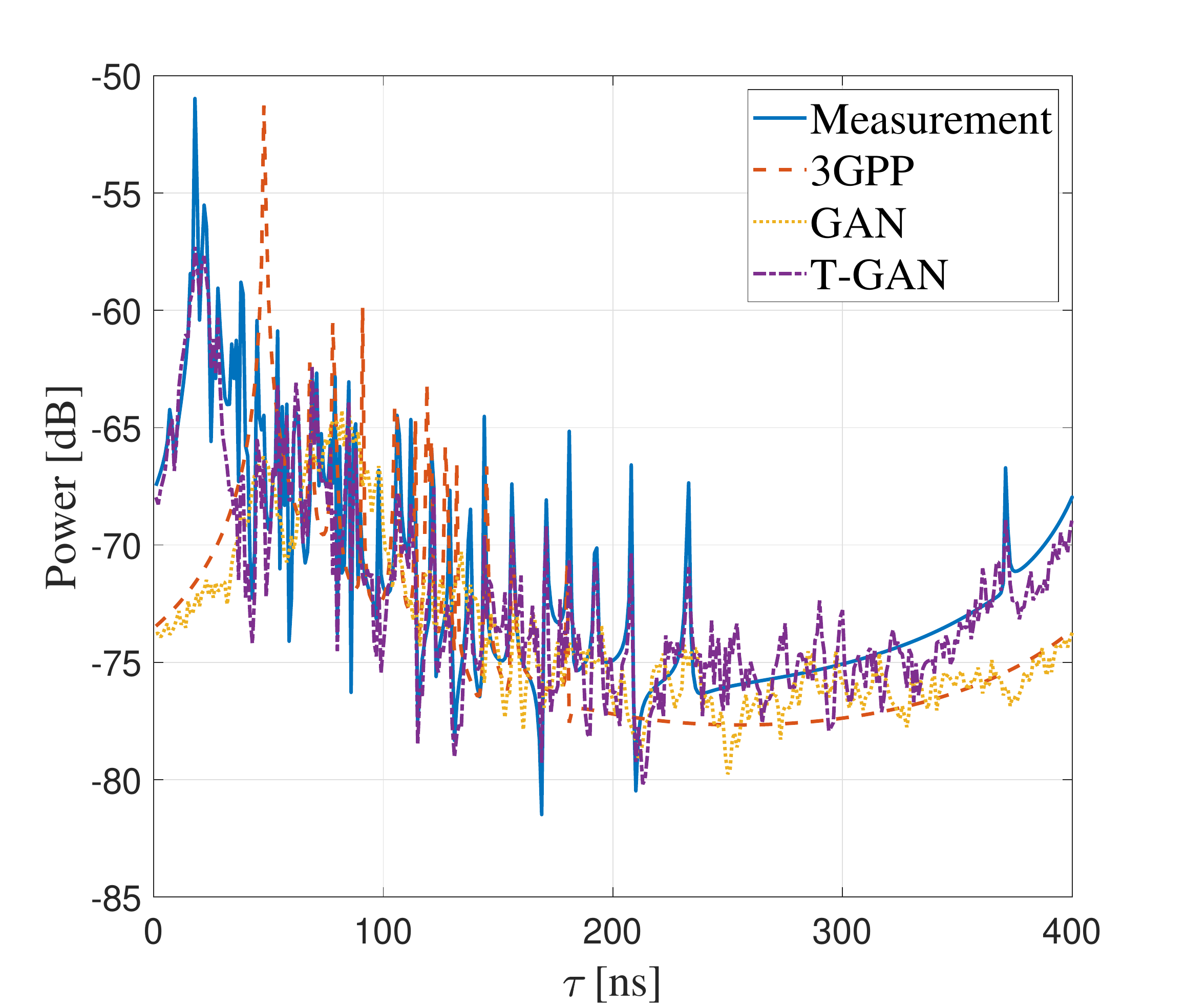}}
    \subfigure[Average PDP.]{\includegraphics[width=0.45\textwidth]{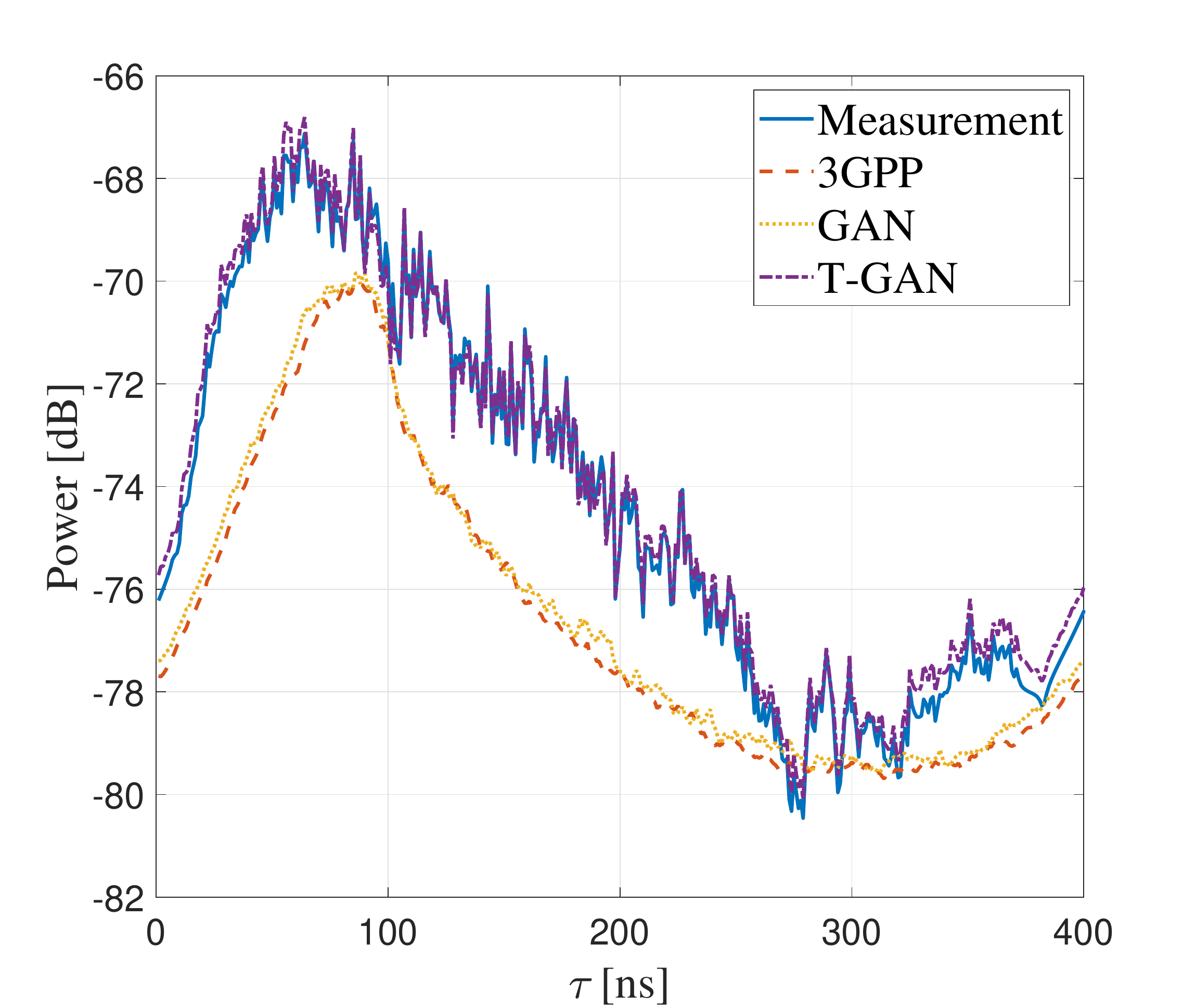}}
    \caption{Plot of PDPs generated by measurement, 3GPP, the proposed GAN and T-GAN.}
    \label{fig_pd}
\end{figure*}

\begin{figure}
    \centering
    \includegraphics[width=0.45\textwidth]{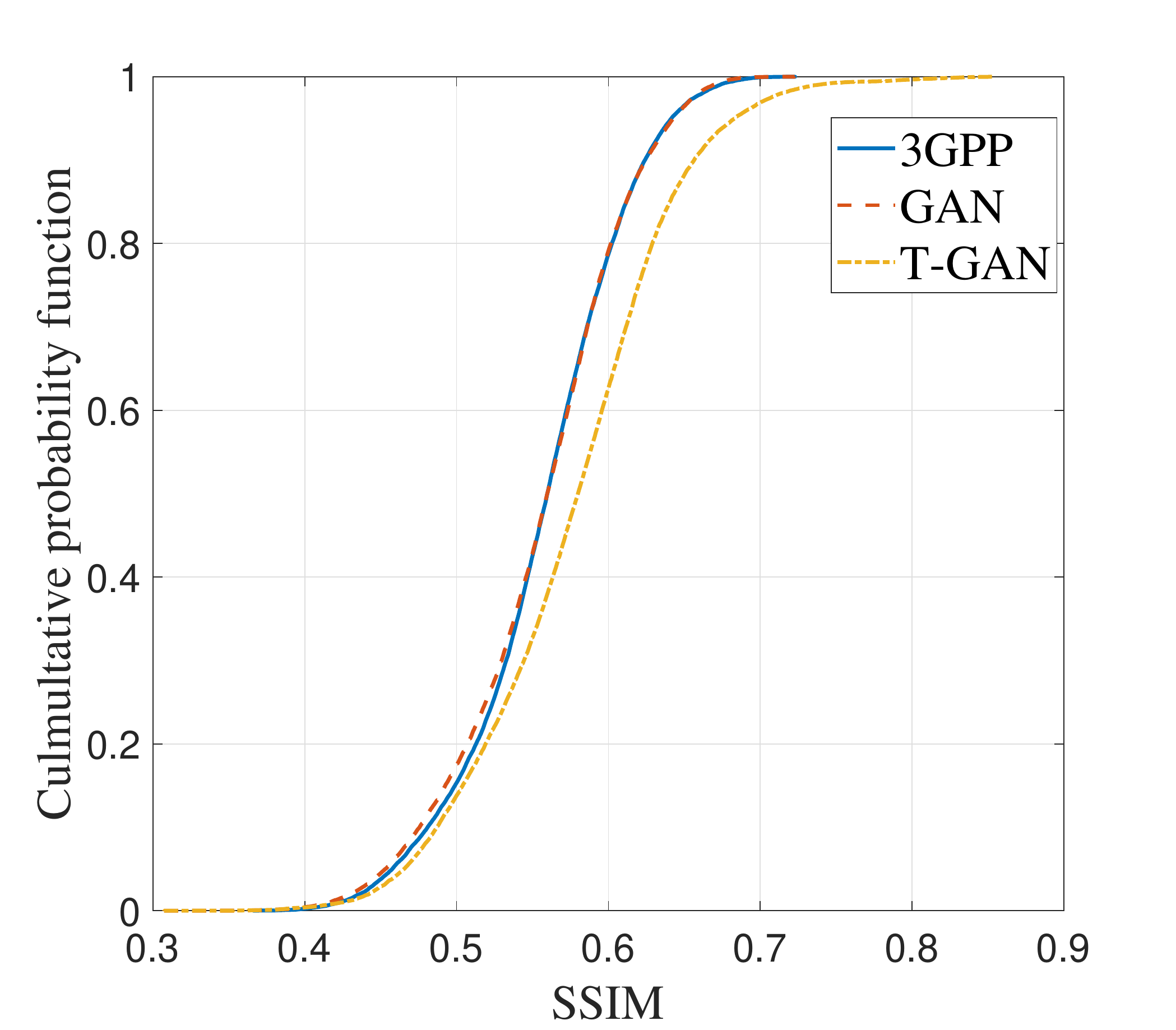}
    \caption{SSIM of PDP for 3GPP, the proposed GAN and T-GAN.}
    \label{fig_cdf}
\end{figure}

\begin{figure}
    \centering
    \includegraphics[width=0.45\textwidth]{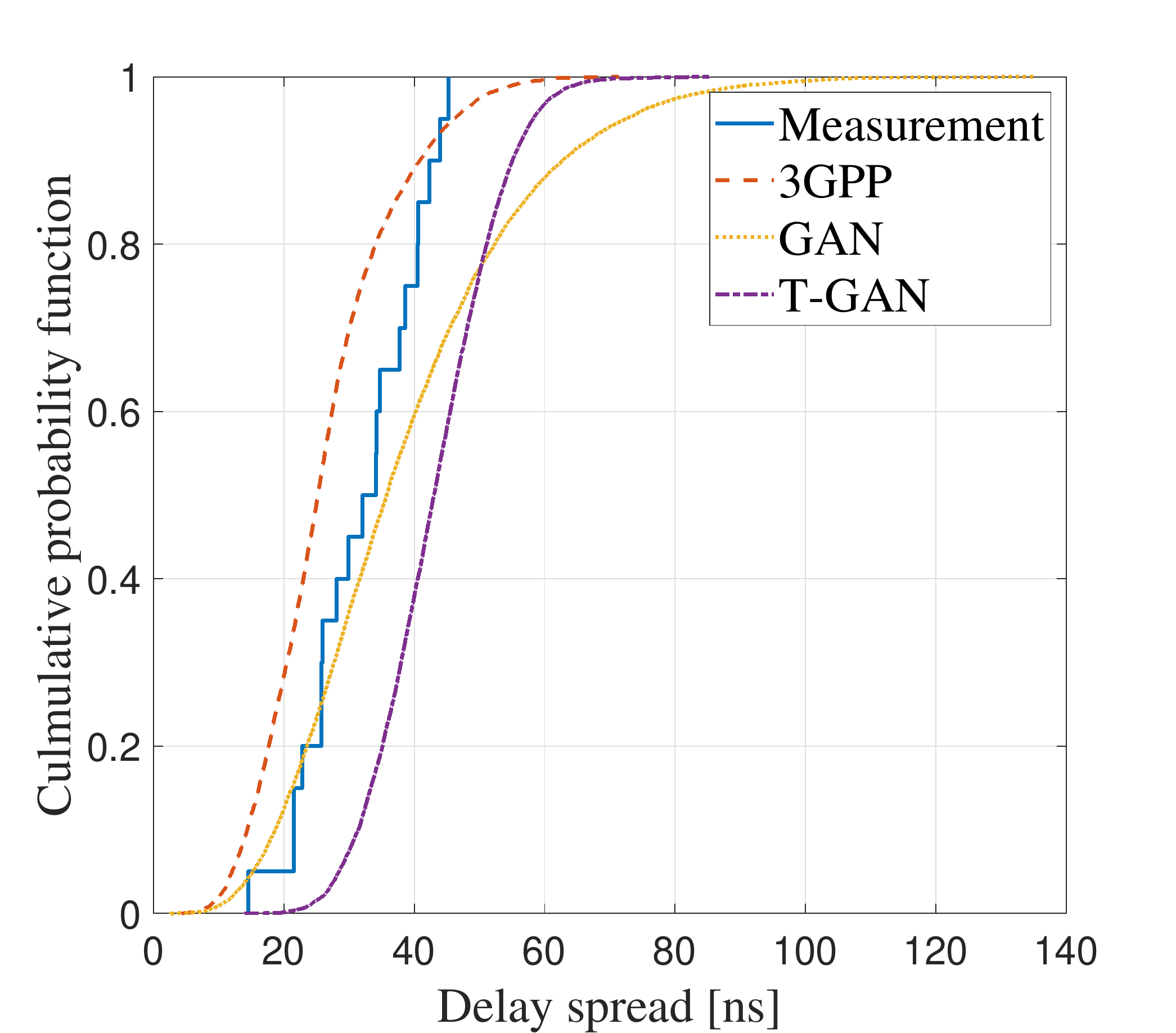}
    \caption{Delay spread for 3GPP, the proposed GAN and T-GAN.}
    \label{fig_delay}
\end{figure}


\subsection{Power Delay Profile}

In the experiment, the samples of PDP from measurement, 3GPP method, the proposed GAN and T-GAN are compared as in Fig.~\ref{fig_pd}(a). It is clear that the PDPs are similar to each other, which proves that the proposed GAN and T-GAN can learn the features of PDPs. Moreover, it is observed that PDP of measurement is more complex than PDP of 3GPP method. There are more peaks and fluctuations in the temporal domain. This shows that 3GPP cannot well capture the channel effects embedded in PDP. Comparing PDPs generated by the proposed GAN and T-GAN, the PDP generated by T-GAN is close to measurement, while the PDP generated  by the proposed GAN is similar to the 3GPP approach. This is reasonable, since the T-GAN can capture the features of PDP from measurement through transfer learning, while the propose GAN can only learn the features of the simulated PDPs by 3GPP method.

In addition, the average PDPs for these method are plotted in Fig.~\ref{fig_pd}(b). It is clear that T-GAN shows good agreement with measurement, while 3GPP and GAN have large deviations from measurement. The deviations can be measured by root-mean-square error (RMSE), calculated as
\begin{equation}
\mathrm{RMSE}=\sqrt{\frac{1}{N_{\tau}}\sum(P_{m}(i\Delta\tau)-P_{g}(i\Delta\tau))^2},
\end{equation}
where $N_{\tau}$ denotes the number of sampling points in PDP, $i$ indexs temporal sample points of PDPs, $N_{\tau}$ represents the number of sampling points and $\Delta \tau$ is the sampling interval. Moreover, $P_m(i\Delta \tau)$ and $P_g(i\Delta \tau)$ are the average power in the $i^\text{th}$ sample point of measured PDPs and generated PDPs, respectively. The results of RMSE for 3GPP, the proposed GAN and T-GAN are 4.29 dB, 4.12 dB and -4.82 dB, respectively. The T-GAN improves the performance of RMSE by about 9 dB, compared with other methods, which demonstrates that the T-GAN outperforms the other methods in terms of modeling the average power of PDP. This is attributed to the powerful capability of GAN in modeling the complex distribution, and the benefits of transfer learning in better utilizing the small measurement dataset.

    

Moreover, to measure the similarity quantitatively,  Structure Similarity Index Measure (SSIM) is introduced, which is widely applied to evaluate the quality and similarity of images. The range of SSIM is from 0 to 1, and the value of SSIM is larger when the similarity between images is higher. The PDPs generated by 3GPP method, the proposed GAN and T-GAN are compared with measurement. The cumulative probability functions (CDFs) of SSIM for these method are shown in Fig.~\ref{fig_cdf}. It can be observed that the proposed T-GAN can achieve higher SSIM values compared with other methods. More than 40$\%$ of SSIM values are higher than 0.6 for T-GAN, compared to only 20$\%$ for 3GPP and the proposed GAN. This further demonstrates the better performance of T-GAN in modeling the PDPs.


\subsection{Delay Spread}

Delay spread characterizes the power dispersion of multi-path components in the temporal domain, which can be calculated as the second central moment of PDPs, by
\begin{equation}
\begin{split}
\Bar{\tau}&=\frac{\sum_{i=0}^{N_{\tau}}i\Delta \tau P(i\Delta \tau)\Delta\tau}{\sum_{i=0}^{N_{\tau}}P(i\Delta \tau)\Delta \tau},\\
    \tau_{rms}& = \sqrt{\frac{\sum_{i=0}^{N_{\tau}}(i\Delta \tau-\Bar{\tau})^2P(i\Delta \tau)\Delta \tau }{\sum_{i=0}^{N_{\tau}}P(i\Delta\tau)\Delta \tau}},
\end{split}
\end{equation}
where $\Bar{\tau}$ denotes the mean delay weighted by the power, $\tau_{rms}$ refers to the root-mean-square (RMS) delay spread, and $P(i\Delta \tau)$ are the power in the $i^\text{th}$ sample point of PDPs.

Then, the CDF plot of delay spread for measurement, 3GPP, the proposed GAN and T-GAN is depicted in Fig.~\ref{fig_delay}. It can be observed that the CDFs of delay spread for 3GPP, the proposed GAN and T-GAN match the measurement well. 

\section{Conclusion}\label{sec_conclusion}
In this paper, we proposed a T-GAN based THz channel modeling method, which can capture the distribution of PDPs for the THz channel with the designed GAN. Moreover, the transfer learning is exploited in T-GAN to reduce the size requirement of training dataset and enhance the performance of GAN, through transferring the knowledge stored in the pre-trained GAN on the simulated dataset to the target T-GAN trained on limited measurement. Finally, we validate the performance of T-GAN with measurement. T-GAN can generate the PDPs that have good agreement with measurement. Compared with conventional methods, T-GAN has better performance in modeling the distribution of PDPs, with 9 dB improved RMSE and higher SSIM. More than 40$\%$ of SSIM values are higher than 0.6 for T-GAN, compared to only 20\% for 3GPP and the proposed GAN.

\bibliographystyle{IEEEtran}
\bibliography{main}
\newpage

\end{document}